\def\be{\begin{equation}}
\def\ee{\end{equation}}
\def\ber{\begin{eqnarray}}
\def\eer{\end{eqnarray}}
\def\bers{\begin{eqnarray*}}
\def\eers{\end{eqnarray*}}

\documentclass[aps,prb,twocolumn,groupedaddress,showpacs,amsmath,amssymb]{revtex4-1}
\usepackage{dcolumn}  
\usepackage{amsmath}
\usepackage{graphicx}
\usepackage{subfigure}
\usepackage{color}
\newcommand{\comment}[1]{}
\usepackage{ifthen}
\newboolean{includefigs}
\setboolean{includefigs}{true}      
\newboolean{includetext}
\setboolean{includetext}{true}     
\newcommand{\condcomment}[2]{\ifthenelse{#1}{#2}{}}

\begin{document}

\title{Low-energy, planar magnetic defects in {BaFe$_2$As$_2$}: \\ $nano$twins, twins, antiphase and domain boundaries}
\author{S. N. Khan,$^{1,2}$ Aftab Alam,$^{3}$ and Duane D. Johnson$^{1,2,4}$}
\email{snkhan@illinois.edu,ddj@ameslab.gov,aftab@phy.iitb.ac.in}
\affiliation{$^{1}$Department of Physics, University of Illinois, Urbana-Champaign, Illinois 61801, USA} 
\affiliation{$^{2}$The Ames Laboratory, US Department of Energy, Ames, Iowa 50011-3020, USA;}
\affiliation{$^{3}$Department of Physics, Indian Institute of Technology, Bombay, Mumbai 400 076, India}
\affiliation{$^{4}$Department of Materials Science \& Engineering, Iowa State University, Ames, Iowa 50011-2300.}

\begin{abstract}
In {BaFe$_2$As$_2$}, structural and magnetic planar defects begin to proliferate below the structural phase transition, affecting descriptions of magnetism and superconductivity.  We study using density-functional theory the stability and magnetic properties of competing antiphase and domain boundaries, twins and isolated $nano$twins (twin nuclei) - spin excitations proposed and/or observed. These nanoscale defects have very low surface energy ($22$-$210$~$m$Jm$^{-2}$), with twins favorable to the mesoscale. Defects exhibit smaller moments confined near their boundaries -- making a uniform-moment picture inappropriate for long-range magnetic order in real samples. {\it{Nano}}twins explain features in measured pair distribution functions, so should be considered when analyzing scattering data. All these defects can be weakly mobile and/or have fluctuations that lower assessed ``ordered'' moments from longer spatial and/or time averaging, and should be considered directly.

\end{abstract} 
\date{\today}
\maketitle

\section{Introduction}
{\par} Fe-based superconductors (FeSCs) provide another avenue to understand  unconventional superconductivity.\cite{RevModPhys.83.1589,AnnualRevCMP.2.121,AnnualRevCMP.1.27,NatPhys.6.645,LowTempPhy.38.888} Due to its ease of synthesis, BaFe$_2$As$_2$ is a prototype for these systems, where its low-temperature (T~$<140$~K) ground state is a striped, antiferromagnetic (AFM) orthorhombic ($Fmmm$) structure,\cite{Rotter08} often called a spin-density wave (SDW), and which is reproduced in Density Functional Theory (DFT) calculations.\cite{Singh08}
At Ne{\'e}l T$_{N}$ ($140$ K), both a magnetic and structural transition occurs to a tetragonal ($I4/mmm$) paramagnet.\cite{Rotter08} By doping with a transition-metal on the Fe-site or others on Ba- and As-sites, superconductivity (SC) can be achieved, and similarly with pressure.\cite{Ni08,Ni10,Alireza09,PhysRevB.78.104527} 

{\par} There are strong connections between the magnetism and SC. Dopants weaken the magnetic state and Cooper pairing is, perhaps, driven by increased magnetic fluctuations out of the ground state.\cite{Kuroki08,Wang09} DFT has proven successful in modeling the geometry, magnetic ordering, and electronic structure of FeSCs. The magnetic ground states of LaFeAsO, BaFe$_2$As$_2$, NaFeAs, and FeTe are all correctly predicted.\cite{PhysRevB.78.085104,Singh08,PhysRevB.80.020504,PhysRevLett.102.177003} Fermi-surface (FS) nesting is apparent from DFT calculations and agrees with angle-resolved photoemission (ARPES), suggesting an itinerant nature\cite{Singh+Du08,Ideta12,Richard10} and which is supported from the spin-wave dispersion.\cite{Harringer11,PhysRevB.78.220501} Furthermore, DFT explains quantitatively effects of doping on FS nesting, and why Cu doping behaves differently than Co and Ni.\cite{Kim12} K$_x$Fe$_2$Se$_2$ (isostructural to BaFe$_2$As$_2$) does not have the hole pockets needed for FS nesting,\cite{Guo10} as  DFT finds.\cite{Cao11} 

{\par} DFT results for BaFe$_2$As$_2$ show a strong coupling between the structure and magnetism.\cite{Singh+Du08,Arturk09} Planar defects, thus, have been proposed to explain key features in magnetic and transport properties of FeSCs near/below the structural transition. Mazin and Johannes\cite{Mazin09} suggested a model in which low-energy magnetic anti-phase (APBs) and 90$^o$ domain (DBs) boundaries proliferate (Fig.~\ref{fig1}), which have yet to be tested.
So, are structural and magnetic planar defects energetically favorable and what are their properties? To answer, we use DFT to model potentially operative magnetic (structural-induced) defects, both isolated and extended, and explore their stability and properties by varying the structural parameters. 

\section{Background}
{\par} Defects can be very important in realistic materials, like BaFe$_2$As$_2$. Above T$_N$, the paramagnetic state may be realized by mobile APBs and DBs; below T$_N$, with interlayer coherence, APBs become pinned and DBs thermodynamically inaccessible, possibly explaining sensitivity to interlayer elements, large magneto-resistance, features in the differential resistivity ($d\rho/dT$), and invariance of resistivity anisotropy.  
With orthorhombic distortions ($a>b$), both structural and concomitant magnetic twins (Fig.~\ref{fig2}) are observed in BaFe$_2$As$_2$ along $\langle110\rangle$ with $100$-$400~nm$\cite{Ma09} up to $10$-$50~\mu m$\cite{Tanatar09} between boundaries.
With stress, samples detwin, but twins return upon its removal;\cite{Tanatar10} as in YBa$_2$Cu$_3$O$_{7-\delta}$,\cite{King1993} $(1\bar10)$ twins terminate on $(110)$ twins.
Twins  cause anisotropic scattering near AFM wavevectors, giving 2-dimensional spin fluctuations. Twins also create stripes of increased diamagnetic response,\cite{Kalisky2010} and nucleate SC at their boundaries.\cite{Xiao2012}
Recently, Niedziela et al.\cite{Niedziela12} found by Rietveld analysis a bigger orthorhombic ratio ({\small O =${(a-b)}/{(a+b)}$}) for local structural fits (O = 1.38\%) than global fits (O = 0.78\%); they proposed a high density of \emph{nano}-twins (Fig.~\ref{fig2}) account for this discrepancy by its better match to measured pair distribution functions (PDF). 
We show that displacements at the \emph{nano}-twin boundary affect spin alignment, reducing the average ``ordered'' moment.  

{\par} For completeness, we note that, while DFT supports the observed SDW for the parent compound, the Fe moment ($1.6-1.9~\mu_B$)\cite{Han09,Singh08} is twice that assessed for the average ordered moment from neutron diffraction ($0.8-1.04~\mu_B$).\cite{Huang08,Wilson09,Gretarsson11} 
In fact, various experiments assess very different Fe moments. Core-electron spectroscopy\cite{Vilmercati12} finds ~$2.1~\mu_B$, like DFT, while $^{57}$Fe M\"ossbauer\cite{Rotter08} and nuclear magnetic resonance\cite{Baek09} find $0.81~\mu_B$, as in diffraction assessments. 
For Fe-based magnets such a large discrepancy between ordered moments from theory and experiment is unusual. 
Spin-orbit and hybridization (controlled by Fe/As planar spacing) in a $\text{DFT+U}$ model explained the small in-plane moments in Fe-pnictides.\cite{Wu08} 
Yet, our DFT moments are reduced $\sim$$10\%$ from spin-orbit, but 50-100\% by slightly reduced Fe-As spacing. 
DFT predicts correct moments at short times ($\sim$10$^{-15}$ s) necessary to yield lattice constants that agree with experiment.\cite{Mazin09}
Dynamical mean field theory (DMFT) explains the discrepancy from DFT as a result of dynamical fluctuations at the Fe sites that reduce the observed moment over longer time scales ($\sim$10$^{-9}$~s),\cite{PhysRevLett.104.197002} and reproduces the trends in reduced Fe moments and renormalized mass across various FeSCs.\cite{Yin2011} DMFT finds FeSCs are correlated due to intra-atomic exchange from Hund's coupling $J$ (0.3\,-\,0.6 eV)\cite{annurev-conmatphys-020911-125045,JPSJ.79.044705,1367-2630-11-2-025021,JPSJ.77.093711} (which reduces the coherence temperature for Fermi liquid behavior\cite{1367-2630-11-2-025021}), not from especially large $U$ (2.8\,-\,5.2 eV, as derived from a five band constrained Random Phase Approximation)\cite{JPSJ.79.044705,JPSJ.77.093711,Anisimov2009442,NatPhys.8.331} or proximity to a Mott insulating state. Below the coherence temperature, high electron mobility results in moment screening (over 10$^{-9} s)$. Notably, this scenario does not consider spatial fluctuations, defects, nor their effect on magnetism near/below the phase transition, as explored in the present work.

\section{Methods: Defects and DFT}
\begin{figure}[t]
\centering
\includegraphics[width=7.0cm]{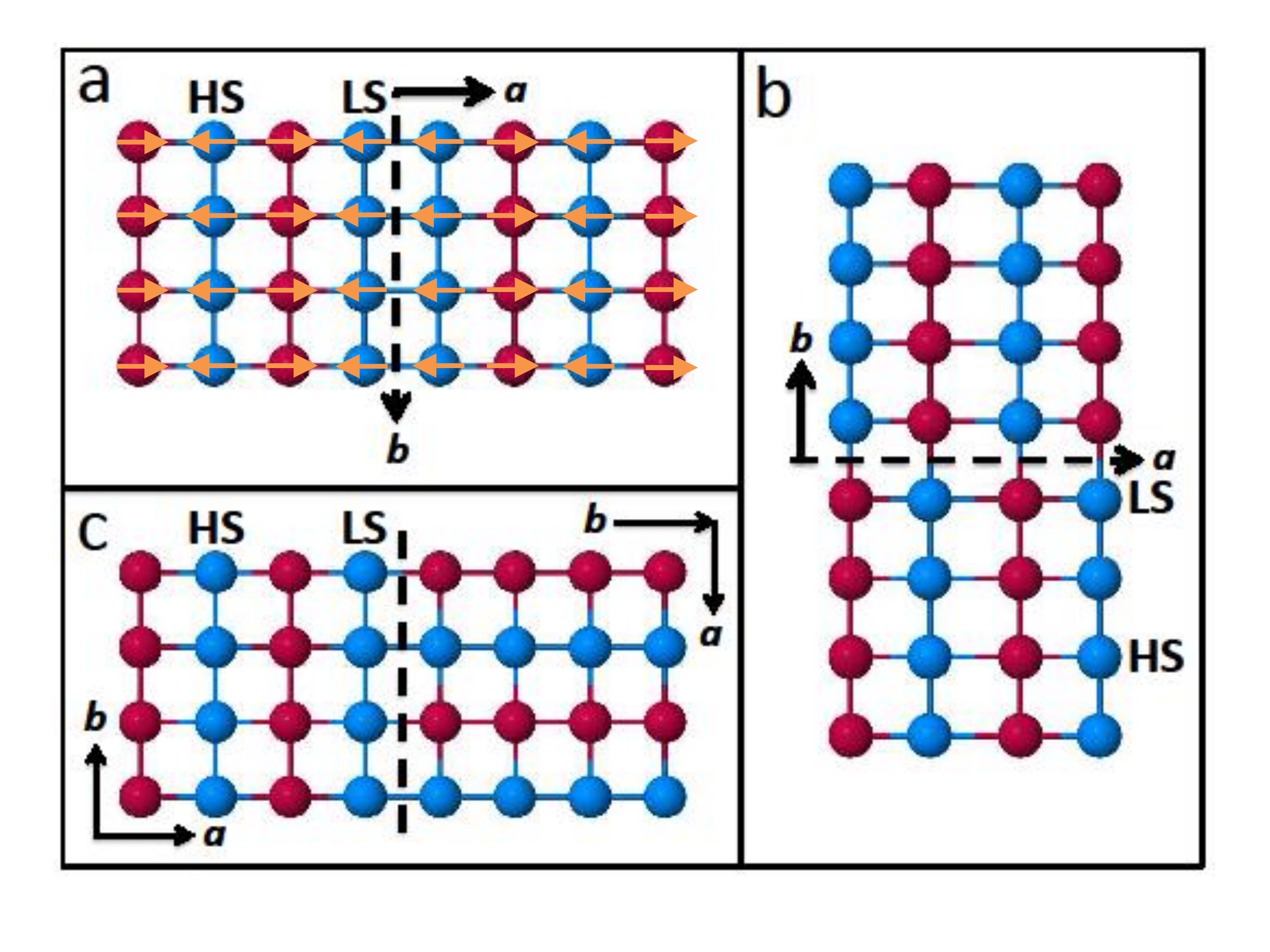}
\vspace{-0.3cm}
\caption {(Color online) APB in the (a) $bc$-, (b) $ac$- planes, and (c) $90^o$ DB with no strain ($a$=$b$). Red (blue) circles are ``up'' (``down'') in-plane moments, as indicated in (a). HS, MS and LS indicate Fe-sites with high-, medium- and low-spin states.} 
\label{fig1}
\end{figure}

\begin{figure}[t]
\centering
\includegraphics[width=8.7cm]{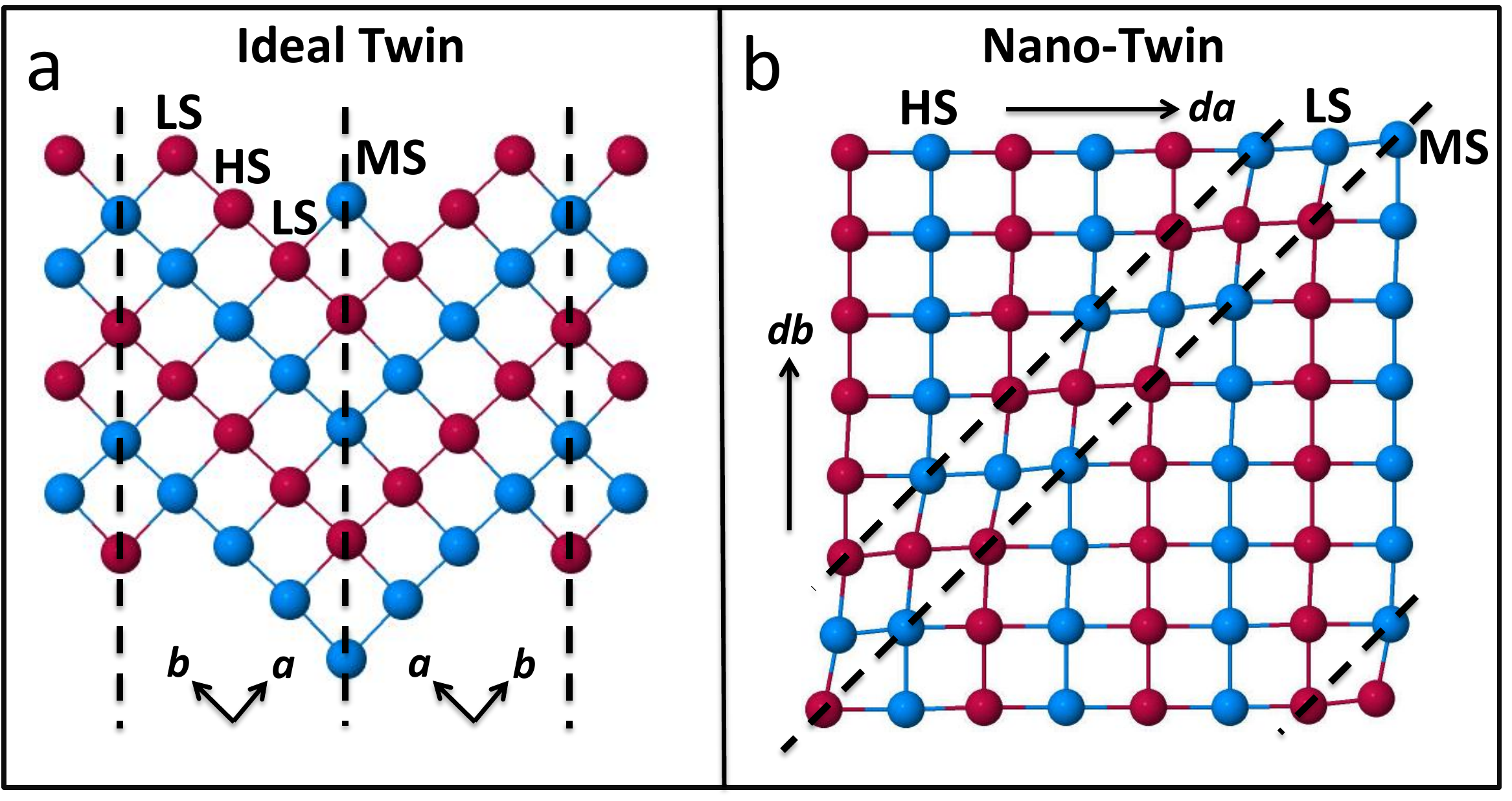}
\caption {(Color online) (a) Twin boundary, without strain, separated by three Fe layers. 
(b) Nanotwin with boundary $\perp$ to (110), where atom positions in a distorted cell are barycentric weights of the cell corners. HS, MS  and LS are indicated. } 
\label{fig2}
\end{figure}

{\par} We use DFT to simulate various magnetic planar defects, $i.e.$, two types of APBs, a $90^o$ DB, twin boundaries, and our modified \emph{nano}-twin, which are all \mbox{low-energy} excitations of the SDW. Figure \ref{fig1}a and \ref{fig1}b shows two APB boundaries in the Fe-plane -- parallel to the $bc$- or $ac$-planes -- and  Fig.~\ref{fig1}c shows a locally unstrained $90^o$ DB. Figure~\ref{fig2}a shows a typical example of an ideal twin. A modified $nano$twin with $2$-dimensional structural distortion (consistent with that suggested by Niedziela \emph{et al.}\cite{Niedziela12}) is shown in Fig. \ref{fig2}b with a series of static displacements along $a$- and $b$-axis in the supercell. The undisplaced nanotwin with $1$-layer of Fe separating defect planes is really a magnetic stacking fault (SF); a nanotwin supercell has very different boundary conditions than a twin, with different far neighbors and distances between defect pairs; indeed, ``ideal twin'' supercells formed with 1-layer separation between defect planes (a high density of SFs)  has local environments like the nanotwin, except that twin has symmetric relaxations governed by the supercell periodicity, whereas the nanotwin has asymmetric, localized distortions to match the PDF. While we show the defect energies are similar, a nanotwin, due to its boundary condition and supercell, may be considered a fluctuating twin nuclei, which can have low-spin Fe-sites unavailable in the ideal twin supercell.

{\par} For nondefected (parent) and defected cells we calculate energy per atom and the associated magnetic moments (bulk is $1.6~\mu_B$). 
From this we derive the planar defect energy, $\gamma$, defined as $\gamma = {(E_{\text{def}} - E_0)d}/{V}$, where $E_{\text{def}}$ and $E_0$ are the total \emph{energy per atom} of the defected and nondefected cell, respectively. $d$ is the distance between defect planes and $V$ is the volume per atom. While the energy per atom is helpful, $\gamma$ is the appropriate comparison for cost of creating the defect interface and its dependence on defect density and defect volume.  Note that 2 defect boundaries are created for twins, hence, $2\gamma$ is appropriate defect energy.

{\par}To do this, we use {VASP}\cite{Kresse96} with plane-wave pseudopotential projected augmented wave (PAW) basis,\cite{Kresse99} with an energy cut-off of $380$--$420~e$V. 
A Monkhorst-pack Brillouin zone integration with a $16^3$ {\bf k}-mesh is used for the SDW ($Fmmm$)  structure. 
Smaller {\bf k}-meshes are used for supercells depending on the length coverage along each axis. 

\begin{figure}[t]
\centering
\includegraphics[width=8.5cm]{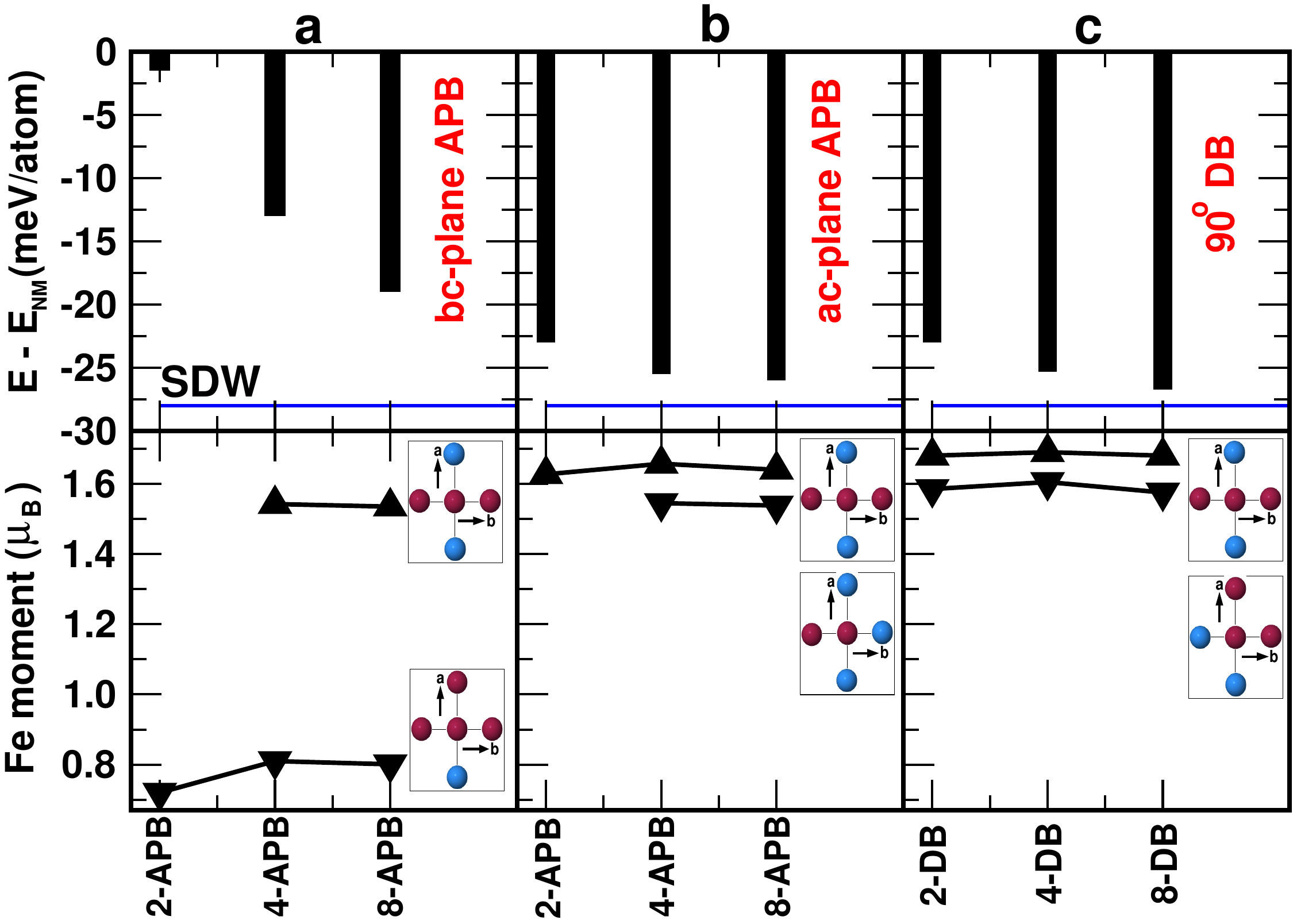}
\caption {(Color online) Energies relative to NM state (top) and Fe moments
 (bottom) for APBs (labels defined in text) in the (a) $bc$-plane (b) $ac$-plane, and (c) $90^o$ DB. 
SDW indicates the AFM ground state. Insets depict the local Fe environment. } 
\label{fig3}
\end{figure}

{\par}For APBs, we constructed doubled ($2$$\times1$$\times1$), quadrupled, and octupled supercells to examine excitations, denoted by 2-APB, 4-APB and 8-APB, respectively (\mbox{Fig. \ref{fig3}}). For  APBs (Figs.~\ref{fig1}a,b), we use measured lattice parameters\cite{Rotter08} ($a$=$5.6146$, $b$=$5.5742$~and $c$=$12.9453$~\r{A}). For a $90^o$ DB, we set $\bar{a}$ = $\bar{b}$ = $(a+b)/2$ = $5.5944$~\r{A} to reduce local strain effects, and construct supercells similar to the APBs, denoted as 2-DB, 4-DB, and 8-DB. Twin ($4[1+n]$$\times2$$\times1$) supercells ($n$=$0,1,...$) are denoted by {($3+4n$)-N} Fe-layers between defect planes, and have $4(1+n)$ unit cells along $a$ and $8(1+n)$$\times10$ atoms/cell. Nanotwin supercells are denoted {3-N}, {5-N}, {9-N}, and {13-N} for Fe-layers between isolated nanotwin  pairs; the supercells with the static displacements suggested by Niedziela \emph{et al.} are more complex because the local distortions  must be compensated within the cell (Fig.~\ref{fig2}b). 


\section{Results}
{\par}The energies and moments for APB and DB defects relative to the non-magnetic (NM) state are shown in Fig.~\ref{fig3} (top), and compared to the AFM ground state (SDW). In all cases, Fe moments have two behaviors: a high-spin state (HS in Fig.~\ref{fig1}) at sites away from boundaries and a low-spin state (LS in Fig. \ref{fig1}) at/near boundaries. For APB($bc$), the LS moment falls substantially to $~0.8~\mu_B$ from $~1.6~\mu_B$, similar to that found by \mbox{Yin \emph{et al.}}\cite{Yin09} While for APB($ac$), the LS moment decreases only to $~1.54~\mu_B$. The two spin states depend on local magnetic environments (inset Fig. \ref{fig3}). Moments do not vary much with the size of the supercells, but these two structures energetically compete with the ground state SDW ($\le 9~me$V/atom). For $90^o$ DB (Fig. \ref{fig1}c), the HS state has a higher moment of $1.7~\mu_B$ due to global strain from changed lattice parameters. The LS moment decreases slightly to $1.57~\mu_B$ near the boundary. This defect requires within $2~me$V/atom excess energy to form compared to the SDW. It is energetically competing with the APB($ac$). Both defects are then expected to be present at the same temperature. The local environment does not play a significant role, suggesting simple models such as counting the number of aligned neighbors is not sufficient to characterize the moments.

\begin{figure}[t]
\centering
\includegraphics[width=8.5cm]{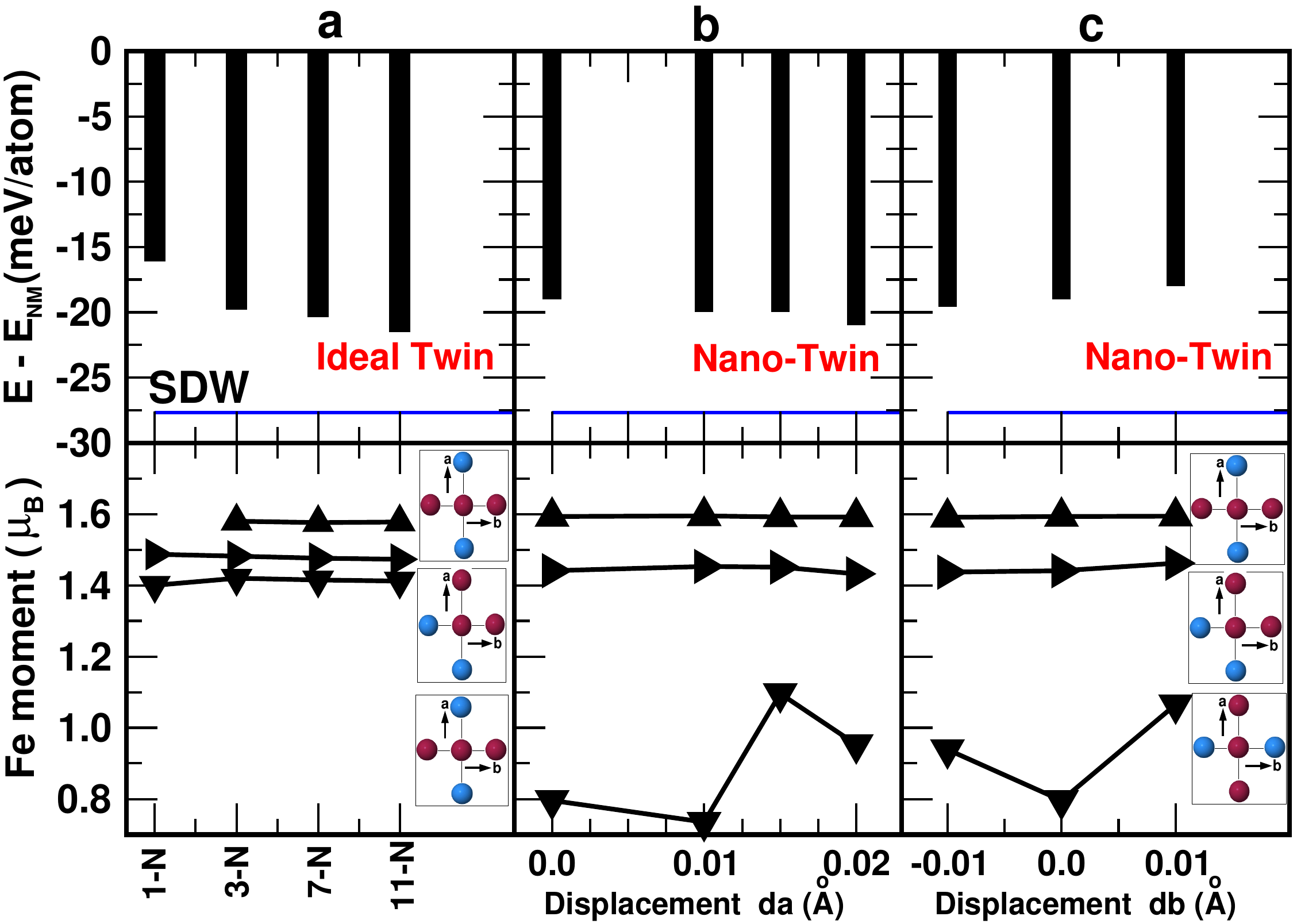}
\caption {(Color online) Same as Fig.~\ref{fig3} but for (a) twins (labels defined in text), and nanotwins ($3$$\times3$$\times2$ supercell) with displacements along (b) d$a$ and (c) d$b$ axes. Fe has HS, MS, and LS states due to local environment (insets).} 
\label{fig4}
\end{figure}

{\par} The energy and moments for twins are shown in \mbox{Fig. \ref{fig4}a.} Interestingly, an Fe-atom in a twin has three spin states depending on the local environment. Fe-atoms at the boundary remain in a medium-spin state (MS in Fig. \ref{fig2}a). A LS state occurs on Fe-sites adjacent to the boundary (Fig. \ref{fig2}a). These Fe-sites have the same nearest-neighbor environment as the bulk HS states but differ in the farther neighbors. These defects can form at a few $me$V/atom, albeit $\gamma$ is more critical, see below.

{\par}The nanotwin energies and moments versus distortion along $a-$ and $b-$axis (in \r{A}) are shown in Fig. \ref{fig4}(b,c).  
Similar to twins, there are three Fe spin states: a HS bulk ($1.6~\mu_B$), a MS ($1.42~\mu_B$) at the boundary, and a LS ($0.8-1.0~\mu_B$) in the vicinity of the distorted side of the boundary. The structural  perturbations show a stronger effect on the LS moments near defect boundaries, decreasing to as low as $0.8~\mu_B$. Isolated (fluctuating) nanotwins are equally competitive to form as dense twins but with much reduced moments. Energies are affected mostly by the changed magnetic configurations and very little by spatial distortions. So, magnetic defects drive the short-range structural distortion (not the other way around) and can help quench magnetization.

\begin{table}[t]
\caption{$\gamma$ ($2\gamma$ for twins) for various planar defects (in $m$J/m$^2$). Energies ($me$V/atom) are relative to SDW.  $\gamma_\text{twin}$ is dominated by $d$ increasing faster than the decrease in ($E_\text{def} - E_0$), unlike for APBs or DBs.}
\label{table1}
\begin{ruledtabular}
\begin{tabular}{lccr}
defect type &  supercell & energy & $(2)\gamma$  \\
\hline
              &       2-APB  &  $26.5$       &     $118$   \\
APB ($bc$-plane)&     4-APB  &  $15.0$       &     $133$   \\
              &       8-APB   &  $~9.0$       &     $160$   \\
\hline
              &       2-APB    &  $~5.0$     &     $~22$   \\
APB ($ac$-plane)&     4-APB    &  $~2.5$     &     $~22$   \\
              &       8-APB    &  $~2.0$     &     $~35$   \\
\hline
              &       2-DB     &  $~5.0$    &     $~22$   \\
$90^o$ DB     &       4-DB     &  $~2.7$    &     $~24$   \\
              &       8-DB     &  $~1.3$    &     $~22$   \\
\hline
``twin'' (ideal)   &   ~$0$-N~   & $18.3$    &     $~57$   \\
``twin'' (relaxed) &   ~$1$-N~   &  $~9.9$    &     $~62$   \\
``twin'' (ideal)   &   ~$1$-N~   & $11.9$    &     $~74$   \\
             &       ~$3$-N~   &  $~8.2$    &     $102$   \\
             &       ~$7$-N~   &  $~6.6$     &    $165$   \\
twin~(ideal) &       ~$11$-N~  &  ${~6.1}$    &     ${228}$   \\
             &       ~$15$-N~  &  ${~5.0}$    &    (max) ${252}$   \\
             &       ~$19$-N~  &  ${~3.7}$    &     ${231}$   \\
             &       ~$23$-N~  &  ${~3.0}$    &     ${222}$   \\
             &       ~$27$-N~  &  ${~2.4}$    &     ${210}$   \\
\hline
  &      ~$3$-N~ &  $~9.2$    &     $~86$   \\
nanotwin             &       ~$5$-N~   &  $~6.4$    &     $~80$   \\
(undistorted)    &       ~$9$-N~   &  $~4.1$     &     $~77$   \\
             &       ~$13$-N~  &  $~2.5$    &     $~63$   \\
\hline
NM bulk      &      $10$ atom &  $28.0$    &     $n/a$   \\
\end{tabular}
\end{ruledtabular}
\end{table}

\begin{figure}[t]
\centering
\includegraphics[width=8.5cm]{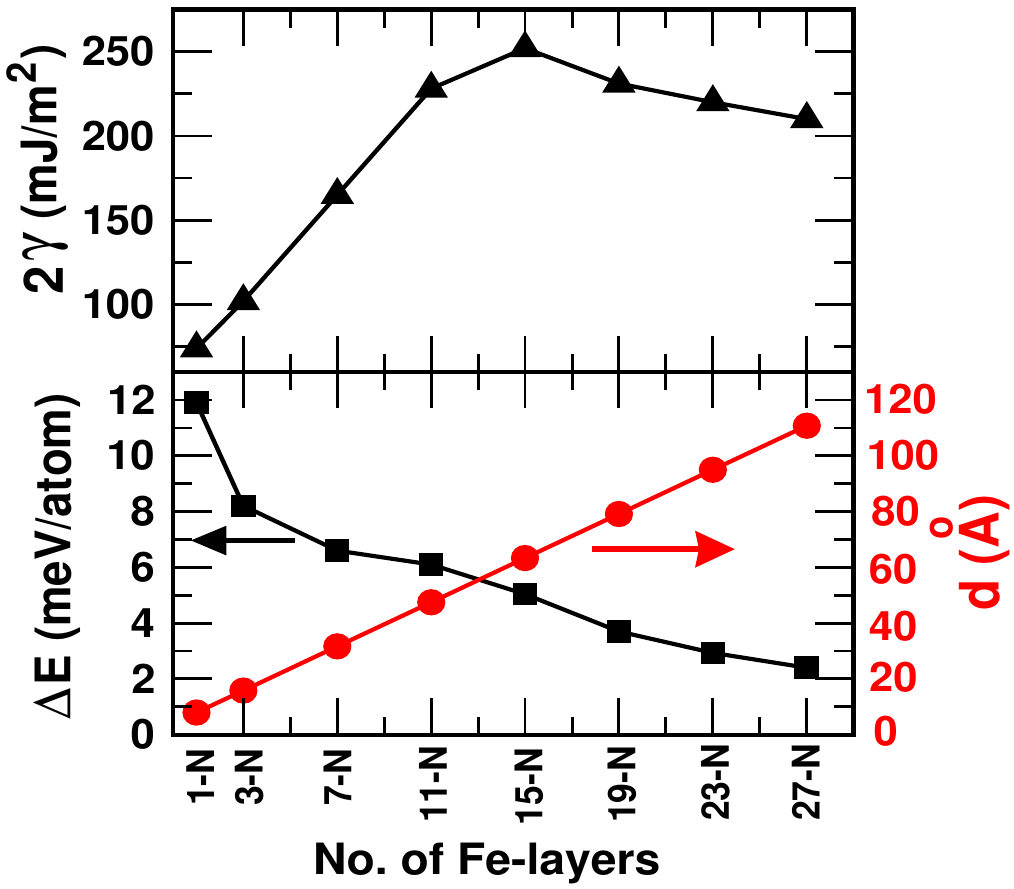}
\caption {Figure 5. (Color online) Twin energy $2\gamma$ (top) and  $\Delta$E = (E$_\text{def}$-E$_0$) and  ``d'' (bottom) versus number of Fe-layers.} 
\label{fig5}
\end{figure}

{\par} Planar defect energies ($\gamma$ or $2\gamma_{twins}$) are compared in Table \ref{table1}; they give the relative {\it order in which magnetic defects can form and remain after processing}. Structural defects can act as pinning sites for magnetic domain walls.
Energetically, APB($ac$), $90^o$ DB and nanotwins (low-energy spin excitations) are the most favorable and most likely to persist after annealing.
Interestingly, densely-pack twins of a single tetragonal variant are also remarkable very low energy.
These nanoscale defects compete with widely separated twins (spin kinks), which are observed. 
Such small fluctuating defects will affect the observed average moments, whereas separated twins will affect the magnetic correlation length, see below.
Separated twins do form and are stabilized by lattice strain arising from disclinations formed when twins oriented $90^o$ apart (from the two tetragonal variants) intersect.\cite{King1993}
It is the twin-twin interactions that stabilize the mesoscale twins.

{\par} Typically in metals, the calculated $2\gamma_\text{twin}$ is monotonically decreasing versus $d$ (the separation of the twin boundaries) until it plateaus at the measured twin boundary energy; essentially, the defects interact (costing energy) until separated enough that they are screened from one another.
Strikingly in BaFe$_2$As$_2$, separated twins are higher in energy than dense twins, until a $d$ of $16$ unit cells ($15$ Fe-layers), where $2\gamma_\text{twin}$ reaches a maximum (Table \ref{table1} and Fig. \ref{fig5}), after which there is a slow convergence of $2\gamma_\text{twin}$ versus $d$ (Fig.~\ref{fig5}). At $28$ unit cells ($\sim$$11~nm$), $2\gamma_\text{twin}$ has not yet converged, {emphasizing the long-range interactions among twins}. Observed structural twins\cite{Tanatar09,Ma09} are extended well beyond the ones computationally feasible. 
Thus, higher-density twins should become prominent near the phase transition, where they compete with the ground state.

\section{Discussion}
{\par} Twin separation $d$ is also affected by stress.
Equilibrium is typically reached when the applied stress is $\sim2\gamma_\text{twin}d$, which is, however,  exceedingly small for isolated twins  in BaFe$_2$As$_2$. 
In real samples twins appear in $90^o$ oriented pairs, where $(1\bar10)$ twins terminate on $(110)$ twins; this configuration is stabilized by lattice strain arising from disclinations,\cite{King1993} where the strain is reduced at the cost of increased $d$. 
With stress (estimated roughly from a set of disclinations,\cite{King1993} and orders of magnitude larger than $2\gamma_\text{twin}d$), samples detwin, but twins would (and do) return upon its removal.\cite{Tanatar10}

{\par} Twins cause anisotropic scattering near AFM wavevectors, giving 2-dimensional spin fluctuations, and create stripes of increased diamagnetic response.\cite{Kalisky2010}
While twin separation depends on local defects and stress, it is expected to get a peak in the magnetic susceptibility $\chi(q)$ at $q=2\pi/{\text{\^d}}$, where ${\text{\^d}}$ is the average twin-twin separation where $2\gamma$ saturates. 
The direction of $q$ is perpendicular to twin boundaries (i.e., $45^o$ to reciprocal-space $k_x$- and $k_y$-axes, where $x$ ($y$) is along $a$- ($b$-) axis). 
While the twins dictate the magnetic correlation length, we suggest that small, low-energy excitation can further depress average moments by spatial and temporal averaging, beyond those due to dynamic fluctuations.

{\par}Nanotwins (Fig.~\ref{fig2}b) with no local distortion are like an isolated, ideal defect pair, not a dense set of twins. 
To understand the effect of short-ranged structural distortion, we have studied a $1$-$\text{N}$ ideal twin with(out) relaxation in $ab$-plane for only those atoms near the boundary, more localized than in the nanotwin supercell. The planar defect energy with(out) relaxation is $62$ ($74$) $m$J/m$^2$. Relaxations along $a$- and $b$-axis lie within $0.9$\% of ideal, close to the best fit to measured PDF,\cite{Niedziela12} so the twin and nanotwin are very similar in energy and local structure. Unlike for ideal twins, the nanotwin surface energy decreases to its limiting value as the nanotwin-nanotwin distance grows (Table \ref{table1}), and it is much lower in energy than extended twins. Thus, a nanotwin may be considered a fluctuating twin nuclei, which has many more LS sites (Fig.~\ref{fig4}) not available in a twin supercell, with moments as low as $0.8~\mu_B$ near the defect, similarly to the assessed values in BaFe$_2$As$_2$. Our calculations support Niedziela et al.'s suggestion\cite{Niedziela12} that nanotwins constitute an important fluctuating excitation in BaFe$_2$As$_2$.

\begin{figure}[t]
\centering
\includegraphics[width=8.7cm]{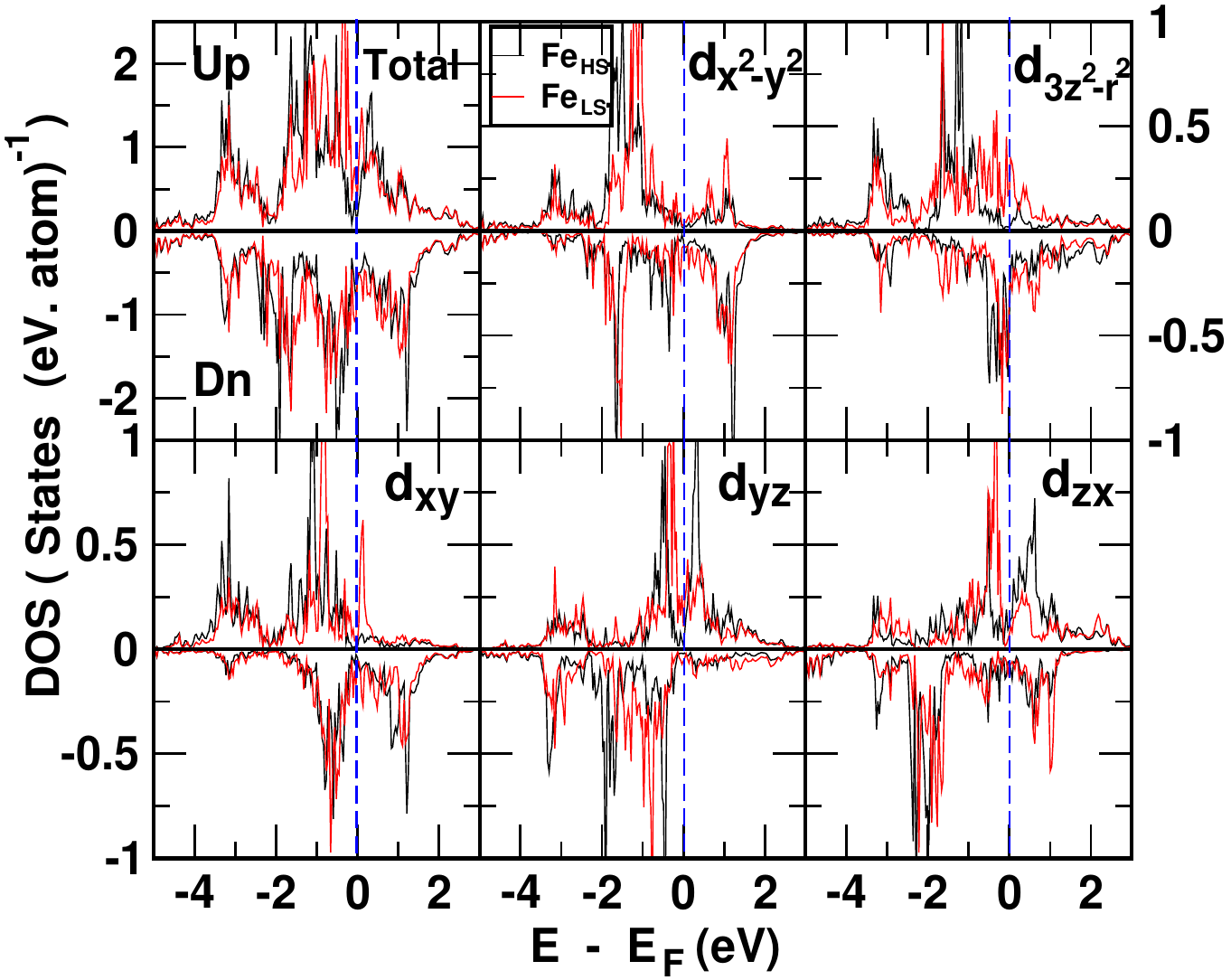}
\caption {(Color online) $d$-orbital and spin-projected DOS of an undistorted nanotwin on HS (black, heavy line) and LS (red, light line) Fe-sites, with majority (minority) DOS plotted on positive (negative) vertical axis. Fermi energy is indicated by vertical dashed line. $x$, $y$, and $z$ directions correspond to $a$, $b$, and $c$ respectively.} 
\label{fig6}
\end{figure}

{\par} Because the local magnetic configurations play the key role in determining the spin states of Fe, we calculated the site- and \mbox{$l$-projected} density of states (DOS) to understand the electronic-structure origin. Figure~\ref{fig6} shows the Fe $d$-projected DOS for HS and LS states. For the bulk (HS) states, the major contribution at Fermi energy E$_\text{F}$ arises from Fe $d_\text{xz}$ and $d_\text{3z$^2$-r$^2$}$, also evidenced from ARPES.\cite{PhysRevLett.104.057002} All the other orbital components exhibit a pseudo-gap near E$_{\text{F}}$.
For LS-Fe compared to HS-Fe, all the projected DOS are shifted towards E$_{\text{F}}$. The most pronounced effect occurs for $d_\text{xz}$ and $d_\text{3z$^2$-r$^2$}$ character, where majority states for LS fall into a pseudogap for $d_\text{xz}$ but are peaked for $d_\text{3z$^2$-r$^2$}$. Although the change of these orbital states is dominated by in-plane Fe-spin configurations, small contributions also arise from the hybridization with As $p_\text{x}$ and $p_\text{y}$ orbitals (out of the Fe-plane), eventually altering the FS. The large difference in the near E$_{\text{F}}$ (majority) DOS between the HS and LS state points to the orbital dependent electronic origin for quenched moments.

\section{Summary}
{\par} In summary, we studied competing low-energy, magnetic planar defects in BaFe$_2$As$_2$. The favorable defects are APB($ac$), $90^o$DB, and nanotwins, but twins (which are observed) are favorable through the mesoscale. The most pronounced reductions in Fe-moment are near the boundaries of APBs($bc$) and nanotwins. We find that isolated closely-spaced twins (twin nuclei) are energetically favorable and correspond to a recently proposed nanotwin suggested to match the pair distribution function from scattering experiment.\cite{Niedziela12} 
Nanotwins are energetically insensitive to microscopic displacements near the boundary, in contrast to sensitivity to the As $z$ coordinate. APBs along \mbox{$bc$-planes} and \mbox{$ac$-planes} are not equally favorable, an anisotropy not anticipated in the Mazin and Johannes model.\cite{Mazin09} These defects can reduce the Fe moment from spatial averaging, an environmental dependence which is not included in DMFT.\cite{PhysRevLett.104.197002,Yin2011} Assessing these defects and their dynamics can affect magnetism, which can be evaluated via Monte Carlo simulations, and which are planned.

Work was supported by the U.S. Department of Energy, Office of Basic Energy Science, Division of Materials Science and Engineering (seed funding), and, for S.N.K. by the Center for Defect Physics, an Energy Frontier Research Center at ORNL. Ames Laboratory is operated for the U.S. DOE by Iowa State University under contract DE-AC02-07CH11358.


%

\end{document}